\documentclass[twocolumn]{emulateapj}
\usepackage{apjfonts}

\usepackage{graphicx} 
\usepackage{natbib}

\shorttitle{The dynamical state of M\,10}
\shortauthors{Beccari et al.}

\slugcomment{Accepted for publication in ``The Astrophysical Journal''}

\begin{document}


\title{The dynamical state of the Globular Cluster M\,10 (NGC 6254)}


\author{G.Beccari\altaffilmark{1},
M. Pasquato\altaffilmark{2},
G. De Marchi\altaffilmark{1},
E.Dalessandro\altaffilmark{3},
M.Trenti\altaffilmark{4}
\and
M.Gill\altaffilmark{5}}


\altaffiltext{1}{ESA, Space Science Department, Keplerlaan 1, 2200 AG Noordwijk, The Netherlands,
gbeccari@rssd.esa.int}
\altaffiltext{2}{Department of Physics, University of Pisa, Largo Bruno Pontecorvo 3, I-56127, Pisa, Italy.}
\altaffiltext{3}{Department of Astronomy, University of Bologna, Ranzani 1, I-40127, Bologna, Italy.}
\altaffiltext{4}{Center for Astrophysics and Space Astronomy, University of Colorado, Boulder, CO, 80309-0389 USA.}
\altaffiltext{5}{Department of Astronomy, University of Maryland, College Park, MD 20742-2421}

\begin{abstract}

Studying the radial variation of the stellar mass function in globular
clusters (GCs) has proved a valuable tool to explore the collisional
dynamics leading to mass segregation and core collapse. Recently,
Pasquato et al. (2009) used the mass segregation profile to investigate
the presence of an intermediate-mass black hole (IMBH) in NGC\,2298.  As
a relaxed cluster with a large core, M\,10 (NGC\,6254) is suitable for a
similar investigation. In order to study the radial  dependence of the
luminosity and mass function of M\,10, we used deep high resolution
archival images obtained with the Advanced Camera for Survey (ACS) on
board the Hubble Space Telescope (HST), reaching out to approximately
the cluster's half-mass radius ($r_{hm}$), combined with deep Wide Field
and Planetary Camera 2 (WFPC2) images that extend our radial coverage to
more than $2 \, r_{hm}$. From our photometry, we derived a radial mass
segregation profile and a global mass function that we compared with those
of simulated clusters containing different energy sources (namely hard binaries and/or
an IMBH) able to halt core collapse and to quench mass segregation. A
set of direct N-body simulations of GCs, with and without an IMBH of mass
$1\%$ of the total cluster mass, comprising different initial mass
functions (IMFs) and primordial binary fractions, was used to predict
the observed mass segregation profile and mass function. 

The mass segregation profile of M\,10 is not compatible with cluster
models without either an IMBH or primordial binaries, as a source of
energy appears to be moderately quenching mass segregation in the cluster.
Unfortunately, the present observational uncertainty on  the binary
fraction in M\,10 does not allow us to confirm the presence of an IMBH in
the cluster, since an IMBH, a dynamically non-negligible binary fraction
($\sim 5\%$), or both can equally well explain the radial dependence of the
cluster mass function.

\end{abstract}


\section{Introduction}

Globular Clusters (GCs) are important astrophysical laboratories for the
study of both stellar evolution and stellar dynamics. In recent years it
has become clear that these two astrophysical processes cannot be studied
independently: physical interactions between single stars as well as the
formation, evolution, survival and  interactions of binary systems have
a significant role in the evolution of the clusters and of their stellar
populations~\citep{spi87}. Such interactions change the  energy budget
of the cluster and therefore influence the time scales on which mass
segregation, core collapse and other dynamical processes occur. On the 
evolutionary side, they can generate objects that cannot be explained by
standard stellar evolution \citep[like blue stragglers, X-ray binaries,
millisecond pulsars, etc.; see][and references therein]{fe06}. 

Recently, various clues have emerged that point to the presence of
intermediate-mass black holes (IMBHs) in GCs. On the theoretical side at
least  three different formation mechanisms have been proposed, namely
merging of stellar-mass black holes via four-body interactions
\citep{mil02}, runaway merging of massive  stars  \citep{por04}, and the
death of Pop III stars in the early universe, which would leave IMBHs
behind as remnants, as shown e.g. by \citet{mad01} \citep[see
also][]{tre07a}. Detailed collisional N-body
simulations~\citep{ba05,tre07b} and theoretical arguments~\citep{he07}
have shown that an IMBH can be at the origin of  a
shallow central cusp in the surface brightness profile (SBP) of GCs that
have been observed in the projected density profile of a few GCs 
by~\citet{noy06, noy07},~\citet{la07}, and \citet{iba09}.  

Line-of-sight velocity studies were undertaken by \citet{bau03a},
\citet{bosch06}, and \citet{chak06} on M\,15,  by
\citet{geb02},~\citet{bau03b},~\citet{geb05} on G\,1, by
\cite{noy08},~\citet{and09},~\citet{sol09} on $\omega$ Cen,  resulting
for now in no undisputed definitive detection. \citet{iba09}  recently
found a  stellar density cusp and a velocity dispersion increase in the
center of the globular cluster M\,54. This could be explained by the
presence of a $\sim 9400 M_{\odot}$ black hole if the cusp stars possess
moderate radial anisotropy. Proper motion studies from Hubble Space
Telescope (HST) data are likely the best way to attack the problem, but
they are technologically challenging and inherently time-consuming,
requiring multi-epoch HST-quality observations.

Numerical simulations by \citet{bau04} have shown that IMBHs 
produce a clear photometric signature in GCs, in terms of a large core to half-mass-radius ratio.
Moreover, \cite{gill08} show that a quenching of mass segregation is
predicted by N-body simulations of clusters harboring an IMBH, resulting
in an observable signature on the radial dependence of the stellar mass
function in GCs, which can be used to either suggest or rule out the presence
of an IMBH. \cite{pas09} provide the first application of this method,
using the observed mass segregation profile to argue against the
presence of an IMBH in NGC\,2298. More generally, studying the radial
dependence of the stellar mass function in collisionally relaxed GCs can
lead to valuable insights on the underlying dynamics of the relaxation
processes, such as stellar evaporation, mass segregation and core
collapse. 

Mass segregation occurs in star clusters as two-body relaxation
processes drive the system towards energy equipartition. 
Heavier stars sink towards the center of the cluster, while the lighter stars 
preferentially live in the outer regions and are more likely to attain 
escape velocity from the system. 
If an additional energy source is present, such as a dynamically significant
population of binaries or an IMBH, the mass segregation and the core
collapse processes are slowed down and the arising gradient of the IMF is less
steep.  Also a significant population of stellar mass black holes can in
principle segregate in the center of the cluster and sustain a large
core \citep[see][]{mac08}, thereby acting as an energy source. This
effect is observed in our N-BODY simulations with a \cite{sal55} IMF (see Section~\ref{MassSeg}), which
produces comparatively more remnants than a \cite{ms79} IMF. 
However, in such simulations this effect typically lasts a few Gyr, because 
stellar mass black holes kick
each other out of the system via dynamical interactions, and is
therefore unlikely to be at work in M\,10, which is approximately $11.8 \pm 1.1$ Gyr old \citep[][]{sala02}.
In any case, a \cite{sal55} IMF likely overestimate the number of large stellar-mass black holes produced in a real GC,
so any effect of such a \emph{dark core} on the long-term dynamics of the cluster is effectively ruled out.

IMBHs and binaries both work towards reducing the degree of mass
segregation attained by a GC at a given (dynamical) age. A realistic
binary fraction can partly mimick the effects of an IMBH, so a detailed
knowledge of the binary fraction allows more robust conclusions either
in favor or against the presence of an IMBH. So far, however, only for a
few GCs has a reliable  binary fraction been estimated with robust
photometric methods based on main  sequence (MS) broadening
\citep[see][and references therein]{sol07,mil08}, so the type of study
that we discuss here is only possible for a handful of clusters. 

M\,10 is a reasonable candidate for harboring an IMBH since it displays a
ratio  of almost $0.4$ \citep{mc05} between the core radius and the half-light radius,
which in dynamically old clusters  is usually interpreted as the
signature of an extra energy source~\citep[see][]{ves94}. In addition, M10 
is an ideal candidate for the mass segregation method described 
and employed for NGC2298 by Pasquato et al. (2009) because, with a mass of about $1.5 \times {10}^5$ $M_{\odot}$ \citep[][]{mc05}, it 
is dynamically relaxed ($\log(t_{h})=8.86 yrs$; Harris 1996) and has not been overly 
influenced by the Galactic tidal field ($r_t/r_{hm}$  = 11.86 and $R_{gc}$ = 4.6 kpc; Harris 1996).  These conditions 
ensure that it has experienced many initial half-mass relaxation times 
and thus has achieved equilibrium with respect to mass segregation.

The structure of the paper is as follows. We present the data analysis
in Sect.~\ref{Observations} and the resulting color-magnitude diagram of
M\,10 Sect.~\ref{sec_cmd}. The radial dependence of the mass and
luminosity function is showed in Sect.~\ref{lumin}, while in
Sect.~\ref{sec_gmf} the global mass function (GMF) of the cluster is
derived.  In Sect.~\ref{MassSeg} we compare the observed mass
segregation profile with predictions from our N-body simulations. A
summary and conclusions follow in Sect.~\ref{Conc}.

\section{Observations and analysis}
\label{Observations}

The photometric data used here consist of two main data sets. The ``ACS
sample" consists of a set of HST images of the core region obtained with
the Advanced Camera for Survey (ACS) through the F606W and F816W filters
(GO-10775; P.I.: A. Sarajedini; see Table~\ref{tab1}).  The ACS Wide
Field Channel (WFC), employs a mosaic of two $4096\times2048$ pixels
CCD, providing a plate-scale of $0\farcs05$/pixel and a total field of
view (FOV) of $3\farcm4 \times 3\farcm4$. The dithering pattern used in
the observations allows us to sample  cluster stars from the very center
out to the nominal half-mass radius ($r_h \sim 1\farcm81$;
\citet{ha96}). Since we were interested in deriving accurate photometry
of the cluster's low-mass MS population ($M\leq0.3M_{\odot}$), we retrieved from the
ESO/ST-ECF Science Archives only images with long exposure times
($t_{\rm exp}=90$\,s). All the flat-fielded images (FLT)  were properly
corrected for geometric distortions and for the effective area of each
pixel, following the prescriptions of \cite{si05}. 

\begin{table}
\caption{Observations}             
\label{tab1}    
\centering                          
\begin{tabular}{c c c c}       
\hline\hline                 
Camera & Filter & \#exp. & Exp. Time \\   
\hline                      
   ACS & F606W & 4 & 90s \\      
   -        & F814W & 4 & 90s  \\
   WFPC2& F606W &  1  & 1100s  \\
   - & F606W &  3  & 1200s  \\   
   - & F814W  & 1   & 1100s \\ 
   - & F814W  & 3   & 1200s \\ 
\hline                                
\end{tabular}
\end{table}

\begin{figure} \centering \includegraphics[width=\textwidth]{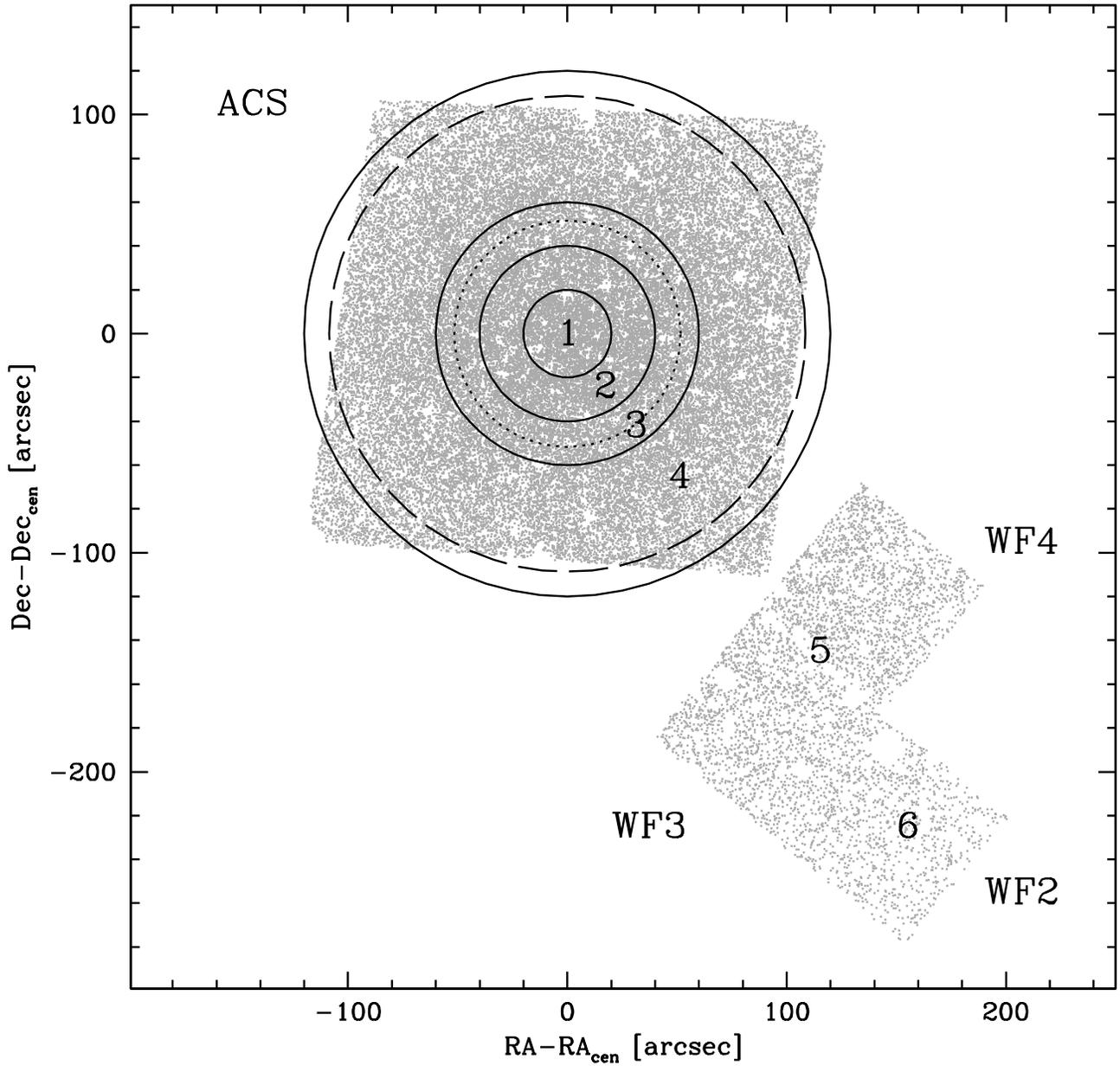}
   \caption{Map of the entire HST data-set used in this work. The
   position of each camera is labeled. The nominal core radius
   ($r_c=0.86\arcmin$) and half mass radius ($r_h=1.81\arcmin$) taken
   from~\cite{ha96} are showed with dotted and dashed circles
   respectively. The full circles define the four annuli in which we
   divided the ACS photometry.  For details on the radial division
   criteria of  ACS and WFPC2 see Section~\ref{star_test}} \label{map}
   \end{figure}

The ``WFPC2 sample" consists of a series of deep HST images obtained with
the F606W and F814W filters of the Wide Field Planetary Camera 2 (WFPC2)
on 1995, September 30 during Cycle 5 (Prop.: 6113; P.I.: Paresce). A
description of the number of exposures and their duration  is given in
Table~\ref{tab1}. The WFPC2 field is located about 3\arcmin\, South-West (SW) of the
cluster center. Also in this case the data were obtained through the
ESO/ST-ECF Science Archives and were pre-processed through the standard
HST pipeline. We did not reduce the PC chip, since  its small FOV 
($\sim 40\arcsec\times40\arcsec$), combined with the extremely low
stellar density at this distance from the cluster center ($\sim 4\arcmin
$), sampled a very small number of objects, with no appreciable
improvement  to the general statistic. A map of the entire data-set and
is shown in Figure~\ref{map}.

\subsection{Data reduction}
\label{riduz}

The goal of this work is to obtain accurate photometry of MS stars down
to the lowest mass allowed by the exposures. Since M\,10 is a reasonably
sparse GC ($c=1.40 $\footnote{Central concentration, $c = log(r_t/r_c)$}; \cite{ha96})  
the images are not affected by severe
crowding, even in the very central regions.  The photometry of two
data-sets was performed as follows. We used DAOPHOTII/ALLSTAR
\citep{ste87} to find stars, derive a spatially varying point spread
function (PSF) and  perform the first measurements for all stars in
every single image. We then used DAOMATCH and  DAOMASTER to  match all
stars in each chip, regardless of the filter, in order to find an
accurate coordinate transformation between the frames.  The  matched
solutions were then fed to MONTAGE2 in order to build a stacked image of
each chip.  In this way we could eliminate all the cosmic rays and
obtain an image of the stars with the highest signal-to-noise ratio.  We
ran the DAOPHOT/FIND routine and the PSF-fitting on the stacked image in
order to obtain a deep master star list.  The master list was then used
as input for ALLFRAME \citep{ste94}, which simultaneously  determines
the brightness for stars in all frames while enforcing one set of
centroids and one transformation between all images.    Finally, all the
magnitudes for each star were normalized to a reference frame and
averaged together, and the photometric error was derived as the standard
deviation of the repeated measures. 

\begin{figure} \centering
   \includegraphics[width=\textwidth]{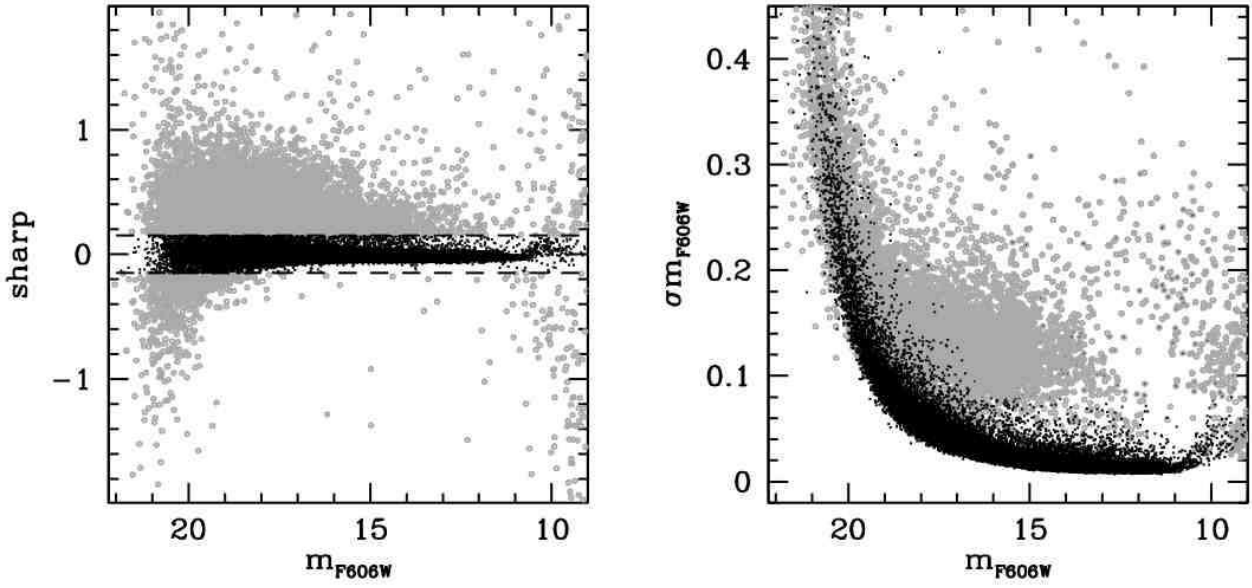} \caption{The sharpness ($sh$) 
parameter provided as an output by ALLFRAME is used
as the source quality diagnostic. By plotting $sh$ as a function of the
magnitude (left panel) we found that spurious objects, detected around the haloes
and along diffraction spikes of saturated stars, have sharpness values
that differ from the ones representative of bona-fide stars ($-0.15 < sh < 0.15$). 
The goodness of the selection  
in sharpness in identifying false detections is confirmed by the fact that the
objects with large $sh$ values have also larger photometric errors
(right panel; see Section~\ref{riduz} for details).} \label{sharp} \end{figure}

We note here that the presence of saturated stars in these deep images
is the major source of artefacts causing spurious detections,
especially in the core regions where most of the red giant stars are
located. In order to clean our catalogues from false detections, we used
a statistical approach similar to the one proposed by \cite{co96}, that
employs the sharpness ($sh$) parameter provided as an output by ALLFRAME
as the source quality diagnostic. By plotting $sh$ as a function of the
magnitude we found that spurious objects,  detected around the haloes
and along diffraction spikes of saturated stars, have sharpness values
that differ from the ones representative of bona-fide stars, which
typically have $-0.15 < sh < 0.15$.   In Figure~\ref{sharp} we show that
the effectiveness goodness of the selection  in sharpness (left panel)
in identifying false detections is confirmed by the fact that the
objects with large sharpness values have also larger photometric errors
(right panel).

We finally transformed the instrumental F606W (V) and F814W (I)
magnitudes to the VEGAMAG system following the prescriptions of
\cite{si05} and~\cite{ho95}\footnote{Notice that we referred to Section
5.1 of WFPC2 Data Handbook for updated values of Photometric Zeropoints}
for the ACS and WFPC2 samples, respectively. The relative star
coordinates of both samples were transformed to the absolute right
ascension and declination values (J2000) using the wide field catalogue
published by~\cite{pol05} as secondary astrometric standard catalogue.

\subsection{The color magnitude diagram}
\label{sec_cmd}

\begin{figure}
   \centering
    \includegraphics[width=\textwidth]{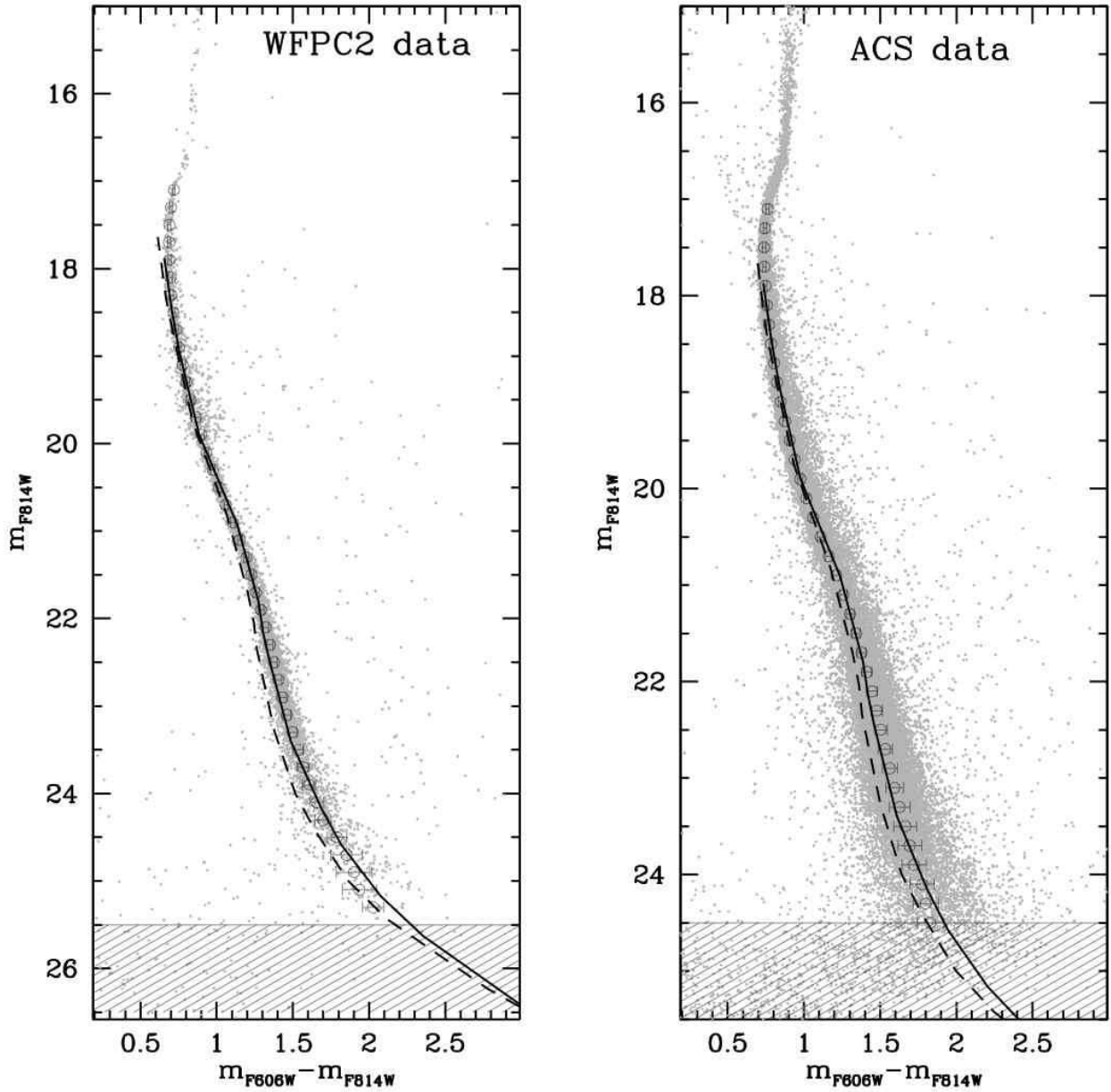}
      \caption{Color magnitude diagram of the cluster as obtained from photometric reduction of the WFPC2 (left panel) and ACS (right panel)  data sets.  The mean ridge line as obtained from 2 sigma rejection fitting process along the MS of each catalogue is also showed (open circles). 
Comparison between the observations and models by~\cite{ba97} for [M/H]=-1.3 (dashed line)
      and [M/H]=-1.0 (full line) is also showed.
The shaded regions in the CMDs show the limits below which the
photometric completeness falls below 50\,\% (see Section~\ref{star_test} for details).}
         \label{cmd}
   \end{figure}

In Figure~\ref{cmd} we show the color--magnitude diagram (CMD) of the
entire data set. The WFPC2 final catalogue (left panel) contains 4995
objects. The same data set have already been reduced by~\cite{dem96}
and  later by~\cite{pio99}. By comparing our photometry with the CMD of
the latter (see their Figure 1), no differences are evident. The
consistency of the two photometric analyses is further confirmed by a
direct comparison of their published luminosity function (LF; see Table
4) and the one calculated in this work for the WF2 chip (see
Section~\ref{lumin} for a detailed discussion).

Due to the long exposure time of the WFPC2 images, the stars at the
turn-off (TO) level in this data set are saturated. In order to obtain a
CMD in the WFPC2 area with a I magnitude range comparable with that of
the ACS data set, we decided to complement our observations with a $V$
vs. $(V-I)$ ground based catalogue from~\cite{ros00}\footnote{The
photometric catalogue can be retrieved at the web page
http://www.astro.unipd.it/globulars/}. This data set was obtained using
the $1.0$\,m Jacobus Kapteyn Telescope (JKT) at La Palma (Canary
Island). We transformed the standard Johnson $V, I$  magnitudes of the
JKT catalogue to  the WFPC2 VEGAMAG system adopting color term as
derived by comparison of all the non-saturated stars in the WFPC2 chips
that were also measured in the ground-based images. No color term is
necessary for the $I$ band. The uncertainties in the $V$ zero points and
in the $(V-I)$ color transformations are  of the order of $0.007$ mag.
Due to the extremely low crowding conditions of the area covered by the
WFPC2 and to the relative proximity of M\,10 ($R_{\rm sun}=4.4$\,kpc;
Harris 1996), the completeness of the ground based catalogue in this
area is expected to be very high. By comparing the observed LFs from the
WFPC2 and JKT data in the same area, we found a good agreement in the
magnitude range $19 < I < 20$ (corresponding to the mass range of 
approximately $0.6M_{\odot}<M<0.7M_{\odot}$). 
We therefore decided to use the JKT
catalogue for stars with $I < 20$ and the WFPC2 catalogue at fainter
magnitudes.

In the right panel of Figure~\ref{cmd} we show the CMD as derived from
the photometric reduction of the ACS data set. The final catalogue
contains 56812 objects. The MS of the cluster is clearly visible and
well sampled well above the TO that occurs at I$\sim17.5$.  A population
of candidate blue stragglers departing from the TO along a bright
extension of the MS is also recognizable. For $I< 16$ saturation starts
to occur, thus making the photometry of  brighter stars unreliable.

According to \cite{pio99}, due to the Galactic latitude of M\,10
($b=23^{\circ}$), the contamination from foreground/background stars is
small and should not effect the LF. On the other hand  \cite{pol05},
using wide field imaging of evolved stellar populations, claim that
field contamination, mainly from the Galactic disk, should be taken into
account. Since no direct measurement of the field contamination in our
magnitude range is possible, we used a statistical approach.  A
catalogue of stars using the Galactic model from~\cite{rob03} covering
an area of 1 square degree around the cluster center was
used\footnote{http://model.obs-besancon.fr/}. After scaling for the ACS
and WFPC2 FOV, we found a contamination of 425 and 159 stars in our
magnitude range respectively for each sample. This means that, even in
the worst case of the WFPC2 sample, where the number of sources is small,
field contamination would be of the order of $\sim 3\,\%$, and thus can
be neglected.

The MS mean ridge lines for the two data sets (open circles in
Figure~\ref{cmd}) were computed by using a 2${\rm nd}$ order polynomial
to fit the locus of MS stars, after rejecting those farther than
$2\,\sigma$ from the best fit line, where $\sigma$ is the combined
photometric uncertainty in $V$ and $I$. Following De Marchi \& Pulone
(2007; hereafter DP07), we applied this same $\sigma-$clipping approach
to identify the bona-fide stars to be used in computing the LF and MF of
the cluster and, in order to increase the statistics, we decided to
accept as bona-fide stars all those within $2.5\,\sigma$ of the ridge
line of best fit. The catalogue obtained after this procedure, 
containing 46\,407 and 4\,390 stars respectively in the ACS and WFPC2
field, will be used hereafter to derive the LF and MF. We mark as shaded
regions in the CMDs of Figure~\ref{cmd} the limits below which the
photometric completeness, as detailed in Section\,2.3, falls below
50\,\%. All the stars in these region are not used for the LF and MF
calculation.

The MS ridge line that we derived in this way agrees very well with the
models of Baraffe et al. (1997). The level of agreement between the
models and the observed data can be fully appreciated in
Figure~\ref{cmd} where the models for metallicity $[M/H]=-1.3$ (dashed
line) and $[M/H]=-1.0$ (solid line) are shown.  As noted by DP07, the
models of~\cite{ba97} are calculated for the F606W and F814W filters of
the WFPC2, which are slightly   different from those on baord the ACS.
We used the same method proposed in their work in order to translate the
filter from the WPFC2 to the ACS system. Briefly, by using synthetic
model atmosphere from e.g. the ATLAS9 library of \cite{ku93}, it is
possible to calculate the magnitude difference in the same filter for
the two cameras as a function of the effective temperature of the star.
As shown in Figure~\ref{cmd} the agreement of the translated model in
the ACS sample (right panel) is very good. The best fit of the model to
the fiducial line corresponds to metallicity $[M/H]=-1.0$, distance
modulus $(m-M)=14.21$ and color excess $E(B-V)=0.26$.
These values are in excellent agreement with those of \cite{pio99}, namely 
$(m-M)=14.20$ and $E(B-V)=0.29$. The metallicity of the best fitting
model, $[M/H]=-1.0$, is fully consistent with the value measured
observationally by \cite{cg97}, namely $[Fe/H]=-1.41$, whereas those for
$[M/H]=-1.3$ deviate in color in the lower part of the CMD, as already
pointed out by \cite{pol05}. We used an isocrone calculated for an age of 10Gyr,
the only available at the moment from~\cite{ba97}. It is clear from their Figure 3a that
the age plays a minor when void effect in the range of masses considered in this work.
In light of the good agreement between the
observed ridge line and the models of~\cite{ba97}, we feel
confident that we can use the latter to translate magnitudes into
masses.

\subsection{Photometric completeness}
\label{star_test}

A crucial step in determining an accurate LF is the estimation of the
photometric completeness of the data. Completeness corrections were 
determined via artificial star tests following the method discussed by
\cite{be02}. We first generated a catalogue of simulated stars with a
$V$ magnitude randomly extracted  from a LF modeled to reproduce the
observed LF in the $V$ band. Then the $I$ magnitude was assigned at
each  sampled $V$ magnitude by interpolating the mean ridge line of the
cluster (see Figure~\ref{cmd}). The coordinates of the simulated stars
where calculated in the coordinate system of the reference frame and
then translated  to each single image using the same matching solutions
found in the data reduction phase (see Section~\ref{riduz}). 

It is crucial to avoid interference between the artificial stars, 
since in that case the output of the experiments would be biased
by artificial crowding not present in the original frame. For this
reason the frames were divided in a grid of cells of fixed width ($\sim
15$ pixel, i.e. more than 5 times larger that the mean FWHM of the stars
in the frames) and only one star was randomly placed in such a box in
each artificial test run. The artificial stars were added to the real
images using the DAOPHOT/ADDSTAR routine. The reduction process was
repeated on the artificial images in exactly the same way as for the
scientific ones and applying the same selection criteria described in
Section~\ref{riduz}. More than 100\,000 stars were eventually simulated
in each chip, for a total of more that 500,000 stars in the entire data
set. The photometric completeness ($f$) was then calculated as the ratio
of the number of stars recovered after the photometric reduction
($N_{out}$) and  the number of simulated stars ($N_{in}$). 
$f$ is expressed as a function of $I$ magnitude.

Due to the strong gradient in stellar density from the center of the
cluster outwards, there is a significant dependence of the photometric
completeness with radial distance, even though M\,10 is a reasonably
sparse GC and the images do not suffer from dramatic crowding.
Nevertheless, in order to take this radial effect into account, we
divided the entire FoV in 6 zones characterized  by a similar
completeness. In Figure~\ref{map} the black full circle show the regions
numbered from 1 to 6. The innermost four regions are included in the ACS
FOV and are concentric annuli all centered on the nominal cluster
center. We notice here that the center used in this work has been
determined from accurate star counts of resolved stars in the central 
regions of the cluster from archival WFPC2 imaging of M\,10 (Prop. 6607,
P.I. F. Ferraro), as explained in a forthcoming paper (Dalessandro et
al. 2010; in preparation). The adopted center of gravity is $\alpha_{\rm
J2000} =16^{\rm h}\,57^{\rm m}\, 8^{\rm s}.92\,,~\delta_{J2000} =
-4^\circ\,05^\prime\, 58\farcs07$, which is in full agreement with the
one of \cite{ha96}.

\begin{figure} \centering \includegraphics[width=\textwidth]{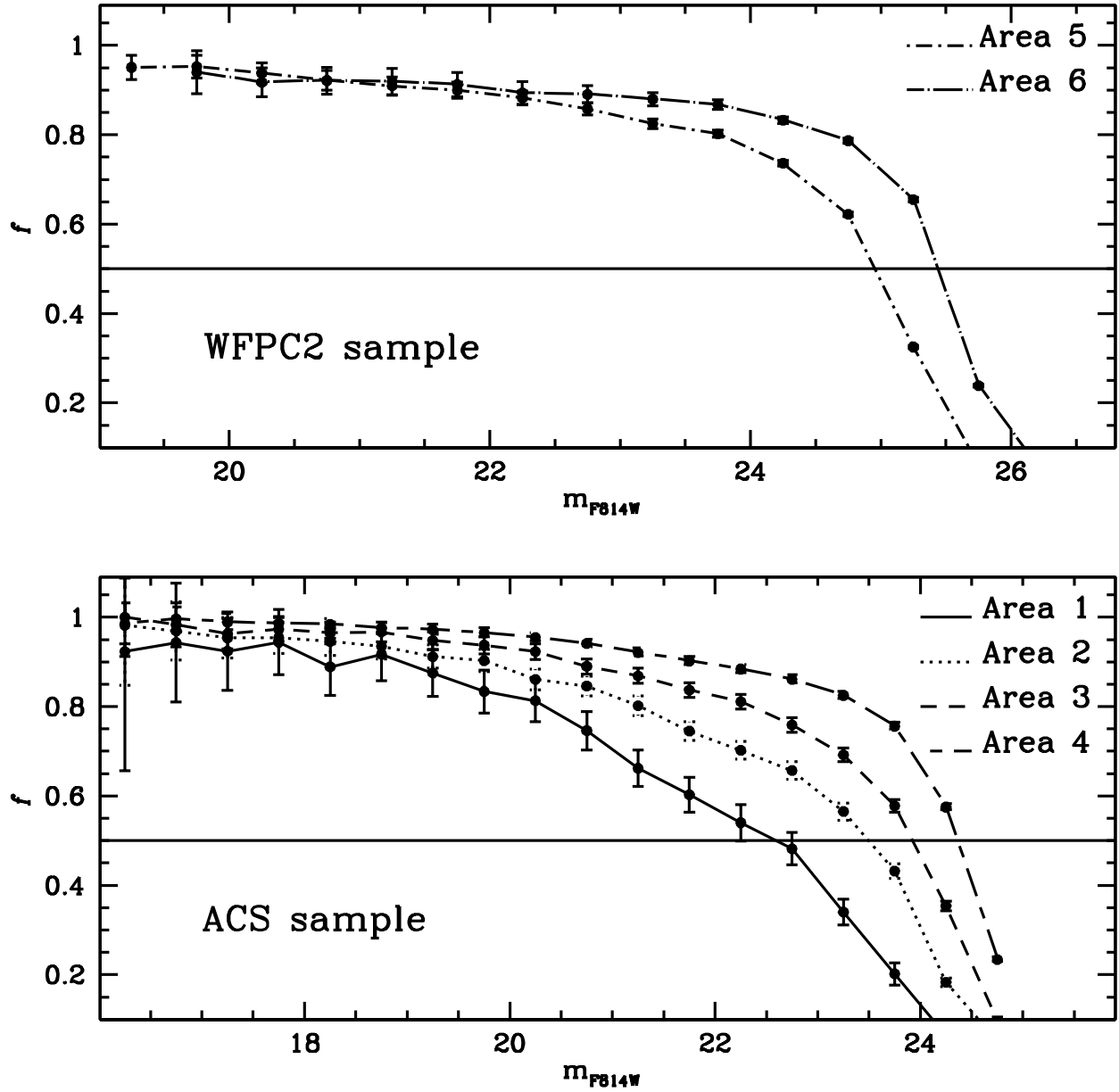}
   \caption{Photometric completeness $f$ as a function of I for the two
   data sets. The ACS and WFPC2 FoV have been divided in 4 and 2 area of
   same completeness respectively. The full horizontal line show the
   limits of 50\% of completeness. } \label{comp} \end{figure}

Region 5 includes both the WF3 and WF4  chips  of the WFPC2,  which
cover and area of equal crowding condition at the same radial distance,
whereas Region 6 corresponds the whole WF2 chip.  In Figure~\ref{comp}
we report the resulting photometric completeness of each of the six
regions as a function of the $I$-band magnitude. It is important to
notice that, even in the very central region of the cluster where 
saturation biases the completeness, artificial star tests show that we
sample the MS stellar population of the cluster down to $\sim 5$ mag 
below the TO (see solid line in lower panel), with a completeness in
excess of 50\,\%.

\section{The luminosity and mass function}
\label{lumin}

\begin{figure} \centering 
   \includegraphics[width=\textwidth]{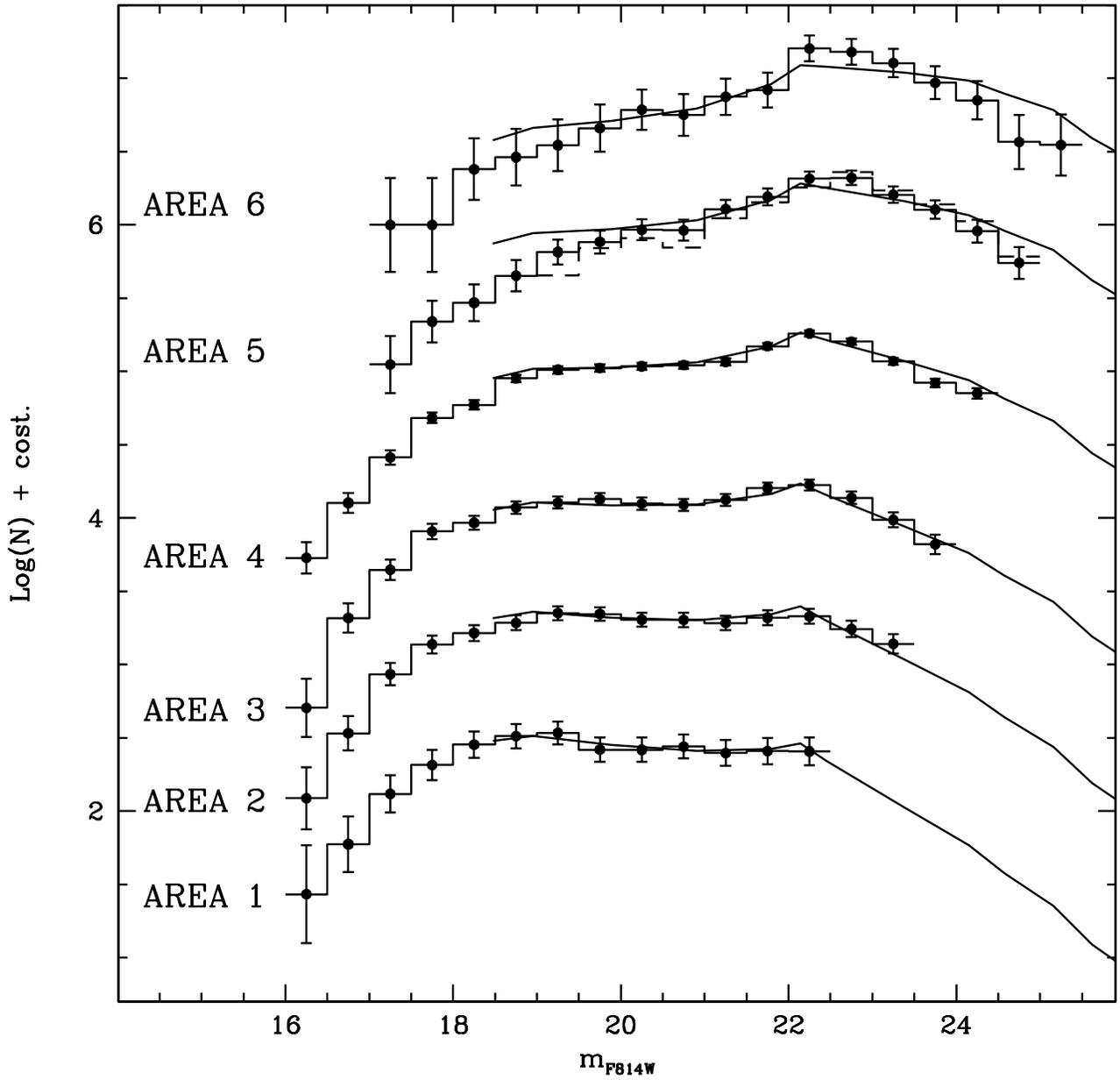}
   \caption{$m_{F814W}$ OLFs of M\,10, divided in different areas (see
   Section~\ref{lumin}). The measurement as reported in Table~\ref{lumcont}
   (full circles in the plot) are shifted by an arbitrary amount to make
   the plot more readable. The theoretical LFs that best fit the data are
   shown as solid lines.  The index of their corresponding power-law MF is,
   from bottom to top, $\alpha=0.7,0.4, 0.1,-0.3, -0.6, -0.9$.  A positive
   index means that the number of stars decreases with mass. All but the
   innermost LF reach a mass of $0.2M_{\odot}$.} \label{lumf} \end{figure}

In order to derive the LF, we selected from our bona-fide photometric
catalogies all stars within each of the six regions mentioned above and
sorted them as a function of the $I$-band magnitude. The resulting
observed LFs (OLFs) are shown as full circles in Figure~\ref{lumf} and
as filled circles after correction for photometric completeness. The
observed and completeness-corrected LFs are also provided in
Table~\ref{lumcont}. In Figure~\ref{lumf} we also show, as a dashed
line, the LF published by~\cite{pio99} in the WF2 frame, which appears
in excellent agreement with ours.

\begin{center}
\begin{table*} 
\scriptsize \tablewidth{12cm}
\caption{Luminosity functions measured in each of the six regions. For 
each region, the table gives, as a function of the $m_{\rm 814}$
magnitude, the number of stars per half-magnitude bin before ($N_{\rm
o}$) and after ($N$) completeness correction and the uncertainty 
($\sigma_{\rm N}$) on $N$.}
\vspace{0.5cm}

\begin{tabular}{c@{\hspace{0.3cm}}
               ccc@{\hspace{0.3cm}}ccc@{\hspace{0.3cm}}ccc@{\hspace{0.3cm}}
               ccc@{\hspace{0.3cm}}ccc@{\hspace{0.3cm}}ccc}
\hline 
  \multicolumn{1}{c}{\,} 
     &  \multicolumn{3}{c}{Area\,1} &  \multicolumn{3}{c}{Area\,2}  
     &  \multicolumn{3}{c}{Area\,3} &  \multicolumn{3}{c}{Area\,4} 
     &  \multicolumn{3}{c}{Area\,5} &  \multicolumn{3}{c}{Area\,6}\\ 
 $m_{\rm 814}$ &  $N_{\rm o}$ & $N$ & $\sigma_{\rm N}$  
     &    $N_{\rm o}$ & $N$ & $\sigma_{\rm N}$ 
     &    $N_{\rm o}$ & $N$ & $\sigma_{\rm N}$ 
     &    $N_{\rm o}$ & $N$ & $\sigma_{\rm N}$
     &    $N_{\rm o}$ & $N$ & $\sigma_{\rm N}$ 
     &    $N_{\rm o}$ & $N$ & $\sigma_{\rm N}$ \\
\hline
17.25   &   121  &   131 &  24 & 259 &  271 &  30 & 269  &   279 &  27 &   511  &   517 &  32 &  28 &     28 &	 7     &    10  &   10  &    4   \\
17.75	&   195  &   207 &  31 & 414 &  434 &  37 & 497  &   511 &  37 &   948  &   961 &  45 &  55 &	  55 &     10  &    10  &   10  &    4   \\
18.25	&   253  &   285 &  38 & 491 &  519 &  41 & 564  &   584 &  39 &  1162  &  1181 &  49 &  74 &	  74 &     12  &    24  &   24  &    6   \\
18.75	&   298  &   325 &  40 & 570 &  610 &  44 & 719  &   744 &  45 &  1745  &  1788 &  63 & 113 &	 113 &     16  &    29  &   29  &    7   \\
19.25	&   299  &   342 &  40 & 647 &  709 &  48 & 763  &   805 &  46 &  1990  &  2047 &  69 & 156 &	 164 &     18  &    35  &   35  &    8   \\
19.75	&   219  &   263 &  33 & 632 &  700 &  47 & 796  &   850 &  47 &  2029  &  2103 &  68 & 183 &	 192 &     19  &    43  &   46  &    9   \\
20.25	&   213  &   262 &  33 & 551 &  640 &  44 & 730  &   791 &  45 &  2065  &  2162 &  68 & 218 &	 232 &     21  &    56  &   61  &   10   \\
20.75	&   206  &   276 &  35 & 540 &  638 &  44 & 691  &   776 &  44 &  2072  &  2202 &  69 & 213 &	 231 &     21  &    52  &   56  &   10   \\
21.25	&   165  &   249 &  35 & 488 &  608 &  44 & 730  &   840 &  48 &  2137  &  2320 &  73 & 293 &	 322 &     26  &    69  &   75  &   11   \\
21.75	&   155  &   257 &  37 & 492 &  660 &  48 & 846  &  1011 &  54 &  2673  &  2963 &  87 & 351 &	 390 &     29  &    76  &   83  &   12   \\
22.25	&   138  &   256 &  41 & 474 &  675 &  50 & 861  &  1062 &  57 &  3204  &  3624 & 101 & 459 &	 520 &     33  &   143  &  160  &   18   \\
22.75	&    ---    &    ---    & --- & 363 &  553 &  46 & 657  &   866 &  52 &  2745  &  3188 &  94 & 451 &	 526 &     33  &   135  &  152  &   16   \\
23.25	&    ---    &    ---    & --- & 247 &  437 &  43 & 424  &   613 &  43 &  1935  &  2345 &  76 & 333 &	 404 &     27  &   112  &  127  &   14   \\
23.75	&    ---    &    ---    & --- &  ---   &  ---   &  ---   & 241  &   417 &  37 &  1260  &  1667 &  65 & 256 &	 319 &     23  &    81  &   93  &   12   \\
24.25	&    ---    &    ---    & --- &  ---   &  ---   &  ---   &  ---   &   ---     &  ---   &   814  &  1416 &  67 & 168 &	 228 &     19  &    59  &   71  &   10   \\
24.75	&    ---    &    ---    & --- &  ---   &  ---   &  ---   &  ---   &   ---     &  ---   &  ---   &  ---   &  &  86 &   138 &  16	 &  29    &  37   &   7	  \\  
25.25	&    ---    &    ---    & --- &  ---   &  ---   &  ---   &  ---   &   ---     &  ---   &  ---   &  ---   &  ---   &  ---   &  --- & ---   & 23    &  35  &     8  \\
\hline
\end{tabular}
\vspace{0.5cm}
\label{lumcont}
\end{table*}
\end{center}

The solid line shown in Figure~\ref{lumf} are the theoretical LFs
(TLFs) obtained by multiplying a simple power-law mass function (MF) of
the type $dN/dm \propto m^\alpha$ by the derivative of the
mass--luminosity relationship of Baraffe et al. (1997) for $[M/H]=-1.0$.
We have adopted a distance modulus $(m-M)=14.21$ and a color excess
$E(B-V)=0.26$ as found in Section~\ref{sec_cmd}. The power-law indices
that correspond to the best fitting models for the six regions, from
Region 1 to Region 6, are respectively $\alpha=0.7,0.4, 0.1,-0.3, -0.6,
-0.9$. With the notation used there, the Salpeter IMF would have 
$\alpha=-2.35$ and a positive index implies that the number of stars is
decreasing with decreasing mass. The wide range of MF slope indices from
the cluster core to $2.5\,r_{\rm hm}$ indicates that stars in M\,10
have experienced the effects of mass segregation. Massive stars are much
more segregated in the central regions were the index is positive. The
MFs are almost flat in the intermediate regions with an inversion of
the  trend outside, where the number of low mass stars is increasing
compared to the more massive ones. We address this issue more
quantitatively in Section\,\ref{sec_gmf}.

\section{The Global Mass Function}
\label{sec_gmf}

In order to further investigate the dynamical state of M\,10, we used
the approach developed by \citet{gu79}. We ran the multi-mass
Michie--King (MK) code originally developed by~\citet{me87,me88} and
modified by~\citet{pu99}, to assess whether the observed MF  variations
with radius are consistent with the degree of mass segregation expected
from energy equipartition due to two-body relaxation. In short, every
model run generates a power-law MF of variable index $\alpha$
characterized by four cluster structural parameters i.e. the core radius
($r_c$), the scale velocity ($v_s$), the central value of the
dimensionless gravitational potential $W_0$ and the anisotropy radius
($r_a$).  The first selection of the models is done by excluding all
those regions in the parameter space that do not fit simultaneously the
observed SBP and the central value of the velocity dispersion
($\sigma_p$). 

The SBP used here was derived from accurate star counts obtained by
combining a photometric catalogue sampling cluster stars in the central
regions from WFPC2 high resoution images and multi-band wide field
images from the  Wide Field Imager (WFI) mounted on the  MPI $2.2$\,m
telescope at La Silla (Chile), extending out to the tidal radius ($r_t =
21\farcm48$; Harris et al. 1996). Details about the SBP can be found in
\citet[][and references therein]{da09}. As regards the velocity
dispersion, the value of $\sigma_p=6.60$\,Km/s comes from \citet{mc05}.

While forcing the MK model to reproduce the SBP and the observed value
of the velocity dispersion will set some constraints on the cluster's
structural parameters and on the range of indices of the MF, in order 
to find a unique solution the additional condition that the model LFs
agree simultaneously with the OLFs at different radii must be imposed
(see e.g. De Marchi et al. 2006; DP07).  This sets very stringent
constraints on the shape of the present  global MF (GMF), i.e. the MF of
the cluster as a whole. 

\begin{figure} \centering 
   \includegraphics[width=\textwidth]{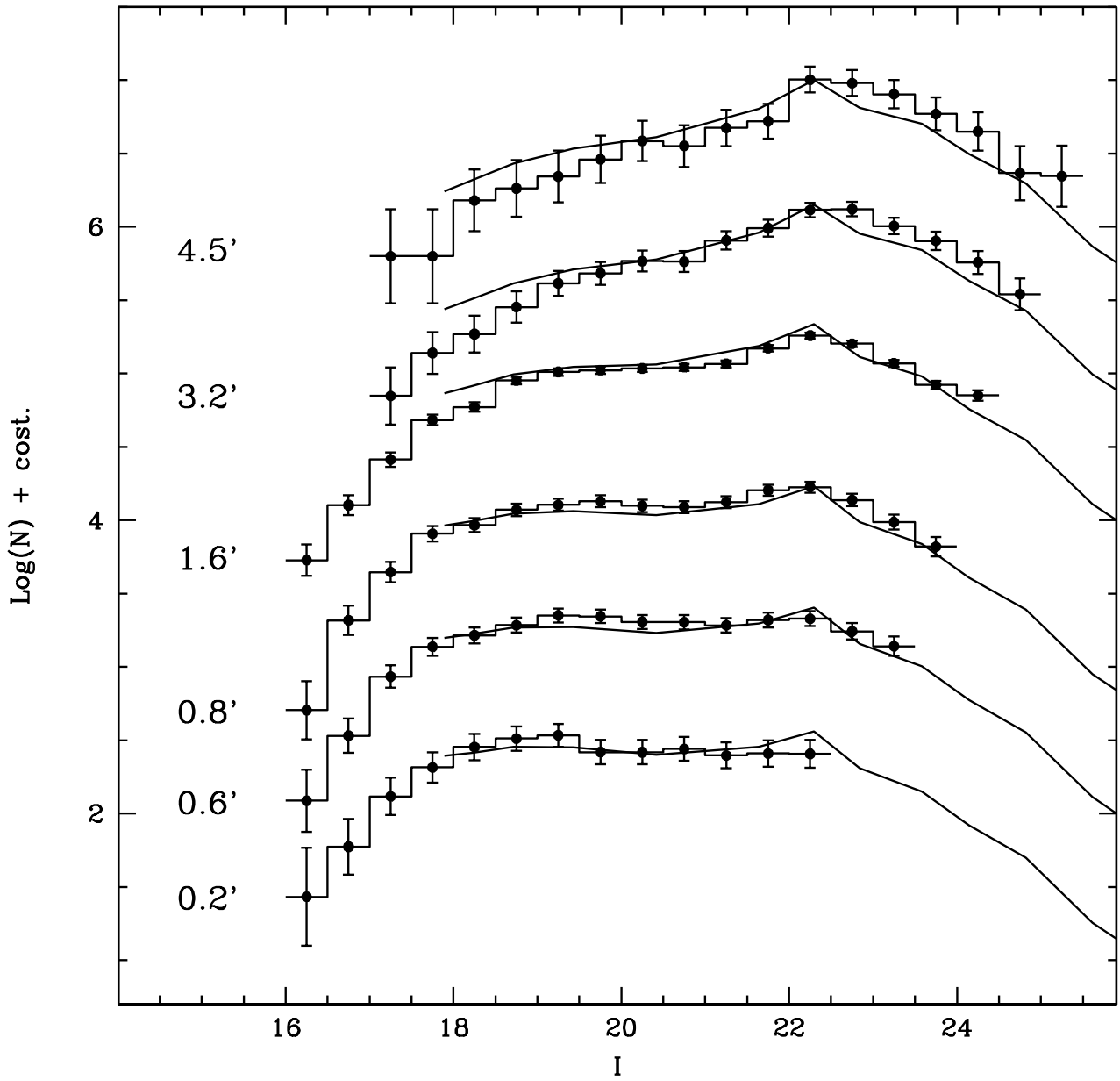}
   \caption{The LFs of Figure~\ref{lumf} (full circles) are compared
   with those predicted by our multi-mass Muchie-King model at various
   radii (reported left of the respective LF) inside the cluster (full line), for a GMF with fixed
   $\alpha=-0.7$. The predicted radial variation of the LF is fully
   consistent with the observations.} \label{mkfl} \end{figure}

We do so in Figure~\ref{mkfl}, where we compare the LFs (corrected for
completeness, solid circles) with the theoretical MFs predicted by the
MK model at the specific radial locations, folded through the M--L
relationship, exactly as we did for the model power-law MFs in
Figure~\ref{lumf}. Here, however, the shape of the local MF is not free,
but is completely set by the two-body relaxation process acting on the
GMF, as per the MK model. The best fitting model also reproduces very
well the SBP with a core radius $r_c=50\farcs4$ and a dimensionless
central potential $W_0=7$, in excellent agreement with the values of 
$r_c=49\farcs41$ and $W_0=6.50$ from~\citet{mc05}.  

The GMF index of the best fitting model is $\alpha =-0.7$ (as already
mentioned in our formalism a Salpeter IMF would have  $\alpha=-2.35$) in
the range $0.25 - 0.8$\,M$_{\odot}$.  In the case of NGC\,2298 (DP07)
and NGC\,6218 (De Marchi et al. 2006), an  excellent match was found
between the index of the GMF and that of the local MF as measured at the
half mass radius. It is  expected that the actual MF near the half mass
radius should be only marginally affected by mass segregation and should
thus be representative of the GMF of the cluster~\citep[see][and
references therein]{dem00}. It is interesting to note that also in the
case of M\,10 the  index of the GMF ($\alpha=-0.7$) is in good agreement
with the value $\alpha\sim-0.6$ (see Section~\ref{lumin}) measured near 
the half mass radius ($r_{hm} = {124}^{\prime\prime}$; see
Section~\ref{MassSeg}).  

The good match between the observed LFs and those from the MK model
suggests that, in its present state, M\,10 is consistent with a
condition of energy  equipartition. On the other hand, since the MK approach
provides by design a snapshot of the cluster evolution, it is not possible to
infer from it the past dynamical history of the cluster, and in
particular whether internal energy sources such as binary stars or an
IMBH are playing any role in the evolution of M\,10. However, this is possible using 
dynamical N-body models that incorporate single and binary star evolution; 
these will be discussed in the next section.

\section{N-body simulations}
\label{MassSeg}

We use state-of-the-art direct N-body simulations to model the dynamics
of M\,10. They are based on NBODY6 \citep{aar03}, modified as discussed
in \citet{tre07b} to improve accuracy in the presence of an IMBH.
Interactions between particles are treated using regularization of close
gravitational encounters with no softening. This choice is motivated by
the need to realistically model the collisional dynamics within the
sphere of influence of the IMBH. An in-depth discussion of the numerical
simulations and the software used is found in \cite{gill08}. We use the
same set of runs presented in that paper, with the addition of further
simulations containing primordial binary fractions equal to $1\%$, $3\%$
and $5\%$. These primordial binaries were assigned binding energies in
the range from $\epsilon_{min}$ to $133 \epsilon_{min}$, where
$\epsilon_{min} = \langle m(0) \rangle {\sigma_c(0)}^2$, $\sigma_c(0)$
is the initial velocity dispersion, and $\langle m(0) \rangle$ the
initial average mass of the system\footnote{ The choice of the number 133 
follows~\cite{he06}, where binary binding energies ranged from $3kT_C$ to $400kT_C$, 
with $T_C$ defined as the central temperature of the system.}. This choice makes our binaries hard,
with separations typically lower than $10$ AU. Binaries are likely to
have a non-negligible impact on the results we obtain, so the addition
of these runs is an important improvement over \cite{pas09}.
The masses of stars in binary systems have been selected independently according to the 
chosen IMF after the instantaneous step of stellar evolution has been applied to evolve the 
turn-off mass to 0.8 $M_{\odot}$. We are essentially adopting a random pairing function 
\citep[see][]{kou09}. Binary exchange interactions are frequent, especially in the 
core of the system, so by $t \gtrsim 5t_{rh(0)}$ (when we consider the simulation snapshots 
for the analysis presented here) more than 50\% of the binaries no longer have their initial companion. 
Hence the details of the pairing function are not expected to significantly affect the measured 
amount of mass segregation.
The initial binary eccentricity distribution is thermal \citep[see][]{tre08}.
For a typical velocity dispersion of 10 km/s, the binary binding-energy distribution 
corresponds to a semiaxis distribution $< 10AU$ \citep[see][]{tre08, tre10}.
Binaries with larger separations would be quickly destroyed by three-body encounters~\citep{he75} 
while binaries with larger binding energy (and smaller semiaxis) than those considered here are generally possible 
only if both components are compact remnants (otherwise their apocenter separation could be less than the physical 
radius of the stars for eccentric orbits). 
In any case such tightly bound objects are essentially dynamically inert 
\citep[see][for runs that consider an extended binding energy range]{tre08}.
Observations of nearby star forming regions and the Galactic field, find a primordial 
binary fraction which is usually quite high \citep[][]{duq1992, kouw2005, kouw2007}
but this refers to all binaries, independent of their binding energy. In our case only
\emph{hard} binaries are included in the primordial binary population, as softer binaries would be quickly
destroyed by three-body interactions, having no long-term dynamical effect. 
This makes our choice of exploring the $0\% - 10\%$ range in primordial binary fraction justified.

Here we summarize the initial conditions to the various runs we use to
explore a variety of different primordial configurations, changing the
IMF to check that our conclusions are independent of the initial
configuration, and the primordial binary fraction to study its influence
on our conclusions. The initial distribution function is a single-mass
\citet{kin66} model, even though a fully realistic mass spectrum is then
used in the N-body calculations. The number of particles is varied from
$N=1.6 \cdot {10}^4$ to $N=3.2 \cdot {10}^4$. If we take the average mass 
of a GC star to be about $0.5 M_{\odot}$, this numbers translate to roughly 
$8 \times {10}^3$ $M_{\odot}$ and $1.6 \times 10^4$ $M_{\odot}$. 
Conversely the mass of M$10$ is estimated by \cite{mc05}
to be $\approx 1.55 \times {10}^5 M_{\odot}$.

None of our simulations includes primordial mass segregation, as all stars start out with the same distribution, regardless of mass. Recently, primordial mass segregation was proposed to explain the extreme depletion of the stellar mass functions observed in several GCs, which exceeds the predictions of primordially non-segregated simulations \citep[][]{bau08}. Primordial mass segregation, in combination with an external tidal field, can be used to justify a much higher rate of depletion of low-mass stars. Moreover, \cite{all09} find that primordial mass segregation naturally arises in clusters initialized in a clumpy, subvirial (dynamically cold) state. In the present work we do not explicitely include mass segregation in the initial conditions. However, in order to draw our conclusions on the presence of an IMBH we consider the mass segregation profiles of relaxed configurations only, by looking at times $t > 5 t_h$. So it is the equilibrium value of mass segregation that matters for our argument, and we expect it not to be influenced by its primordial value, similarly to what is found for $r_c/r_{hm}$, which settles to an equilibrium value after several relaxation times, irrespective of its starting value \citep[see][]{tre07c,tre10}.

Initial stellar masses come from either a \citet{sal55} or \citet{ms79}
IMF. 
We have adopted a lower limit to the IMF of $0.2 M_{\odot}$ and an upper limit of $100 M_{\odot}$ 
(prior to the stellar evolution step). The choice of the lower mass limit is driven by the mass of
the faintest stars that are observable in our data set with a satisfactory completeness ($\approx 50\%$). 
\cite{pas09} computed control runs including stars with masses down to $0.1 M_{\odot}$, 
obtaining the same results as in the case of runs with masses truncated at $0.2 M_{\odot}$.
Stellar evolution is carried out in a single step to a turn-off
mass of $M_{T.O.}  = 0.8 M_{\sun}$, at the beginning of the simulation.
The details and the motivation of this choice are explained by
\cite{gill08}. Half of the runs included in \cite{pas09}, contained
primordial binaries, with a binary fraction of $10\%$. 
This includes runs containing both an IMBH and a $10\%$ binary
fraction, which allow us to study the combined effect of the two
ingredients. Qualitatively, the quenching effects of binaries and of
an IMBH sum, leading to an enhanced depression of mass segregation.
This fraction can reasonably be considered as  the value 
representative of the binary
fraction of a galactic globular cluster~\citep[see][]{be02}.  In this
paper, new simulations without an IMBH but with primordial binary
fractions of $1\%$, $3\%$ and $5\%$ are considered. These new runs have
a \citet{ms79} IMF and $3.2 \cdot {10}^4$ particles. They are included
to give a more fine-grained scale in binary fraction, to properly
compare its dynamical effects with the observations.

In about half of the simulations, an IMBH with mass $M_{IMBH}\simeq
0.01$, or $\sim 1 \%$ of the entire cluster is present. 
Values of the $M_{IMBH}/M_{GC}$ in the range $0.5\%$ to $2\%$ are expected 
by extending the~\cite{mag96} scaling law to GCs. As shown by~\cite{gill08}, 
even rising the mass of the IMBH up to $3\%$ of the cluster mass does not change the amount 
of mass segregation found in the simulations. \cite{gill08} further suggest that, in general, the dependence of the quenching of mass segregation on IMBH mass is weak, given that the quenching mechanism is based on the scattering of stars by the binary system formed in the GC center by the IMBH and a massive companion (e.g. a stellar-mass black hole).
The tidal field from the parent galaxy is taken into account in the simulations
assuming circular orbits with the tidal cut-off radius self-consistent
with the initial cluster concentration parameter. 
 \cite{dine99} find that M$10$ has a moderately elliptic orbit around 
the galactic center, with eccentricity $e = 0.19 \pm 0.05$. This low value of 
eccentricity makes the circular orbits adopted in this work a satisfactory 
approximation to the orbit of M$10$.
For further details of the tidal force treatment see \citet{tre07c}.

\subsection{Comparison with the observations}
\label{compa}

We applied the same algorithm used by \cite{pas09} on NGC $2298$ to the
catalogue of stars we obtained for M\,10, in order to produce a radial
mass segregation profile. A half-mass radius was obtained for main
sequence stars (with masses ranging from $0.26$ to $0.8$ $M_{\odot}$),
by computing their total mass within the radial range covered by the
observations and finding the radius which contains half of it in
projection. A non-parametric mass growth-curve was constructed via
spline interpolation as in \cite{pas09}. We obtain $r_{hm} =
{124}^{\prime\prime}$, which is slightly larger than the half-light
value of ${108.6}^{\prime\prime}$ determined by \cite{ha96}. This is not
surprising and is per se an indication of mass segregation, since bright
giants (with masses around the turn-off mass) dominate the luminosity
profile. We used the obtained half-mass radius to normalize our
mass-segregation indicator, i.e. the mean mass of MS stars as a function
of radius, following \cite{pas09}. The mass segregation profile $\Delta m (r)$ is 
constructed by plotting the mean mass of a main-sequence star 
(normalized by subtracting its value at the half light radius) as a function of radius.

We compared the observational mass segregation profile to the confidence
intervals predicted by means of the set of simulations used in
\cite{pas09} with the addition of the three new IMBH-free runs with a
$1\%$, $3\%$, and $5\%$ primordial binary fraction, consistently
defining the range of observable MS stars from $0.26$ to $0.8$
$M_{\odot}$. \cite{gill08} show that the simulations reach a stationary
configuration with respect to mass segregation after approximately $5$
relaxation times. For the sake of comparing them to the observed mass
segregation profile of M\,10, the simulations were analyzed between $7$
and $9$ initial-half-mass relaxation times.

\begin{figure} \centering
   \includegraphics[width=0.49\textwidth]{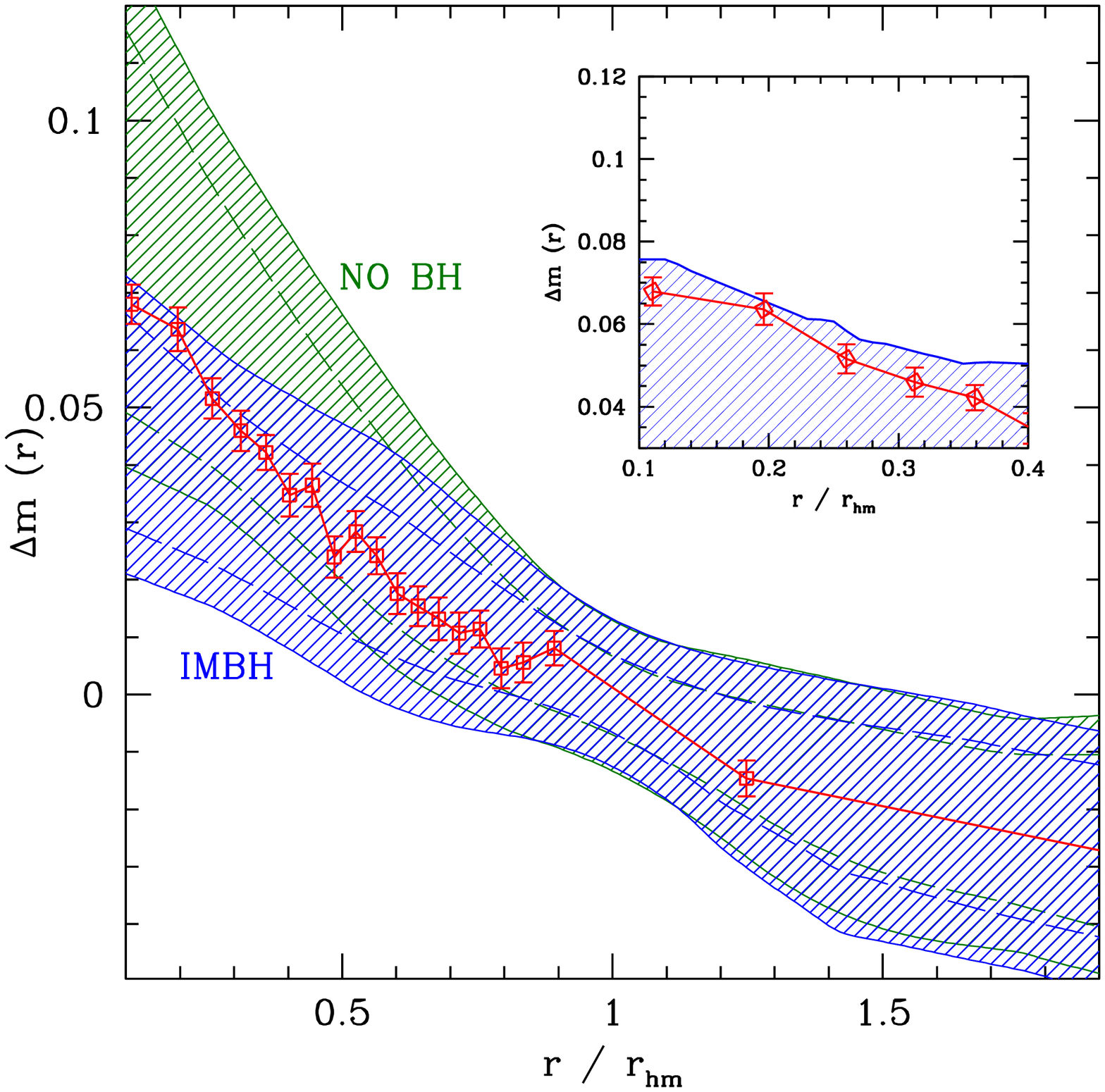}
   \includegraphics[width=0.49\textwidth]{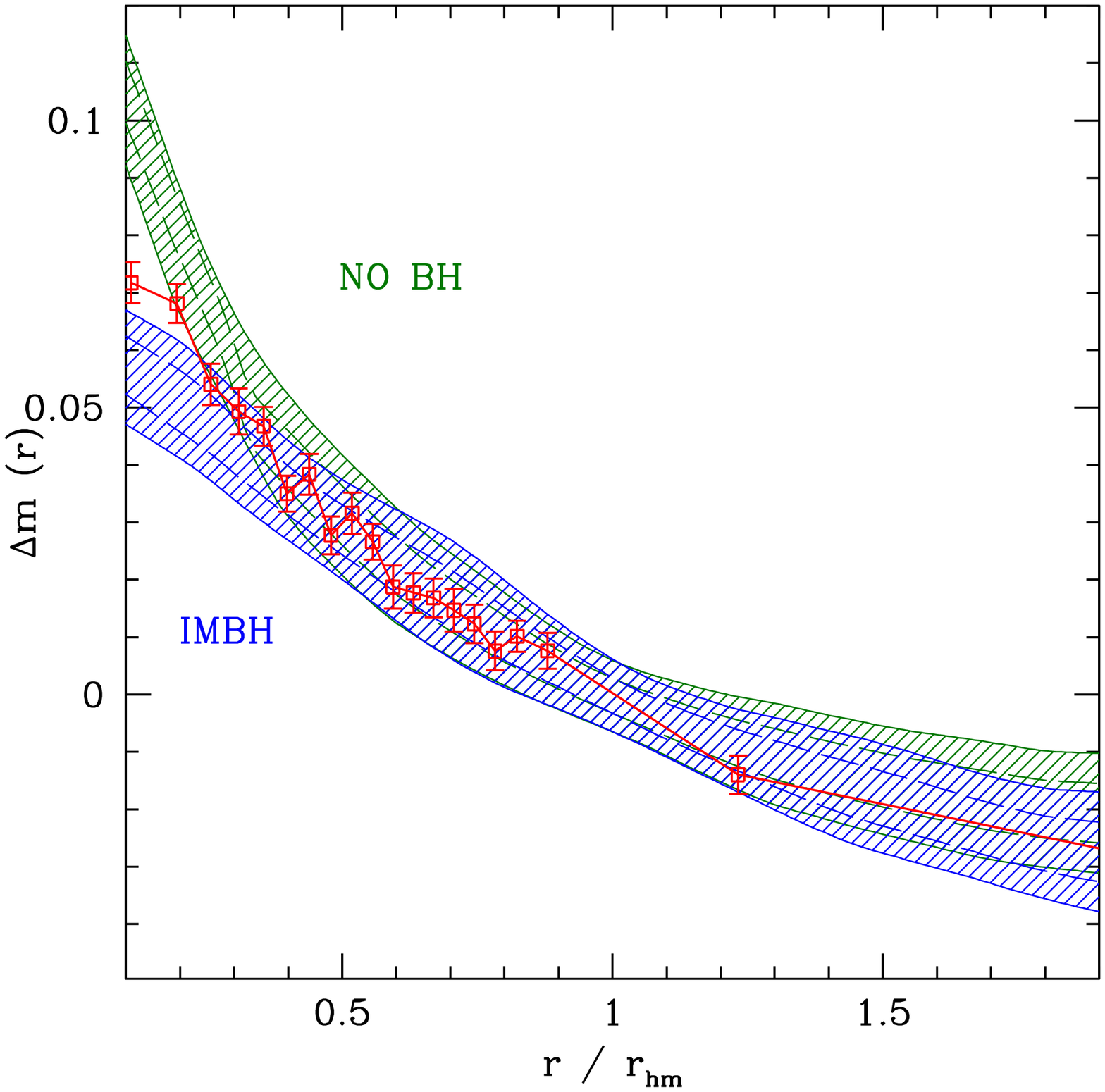}
   \caption{Observed radial mass segregation profile ($\Delta m(r)$
   measured in $M_{\odot}$) for M\,10 (red points with 1$\sigma$ error
   bars), compared to expectations from numerical simulations. 
   The shaded areas in both panels represent the $2 \sigma$ 
   confidence area for our simulations with an IMBH (blue) and without an IMBH (green). 
   The long dashed lines define the $1 \sigma$ confidence regions. 
   The small inset shows the upper envelope of all snapshots of 
   the simulations including an IMBH in the set.   
   On the left panel the simulations contains runs with up to $10\%$ 
   primordial binaries as in~\citet{pas09}. 
   On the right panel only the simulations with Miller \& Scalo IMF, 32k particles and 
   no primordial binaries are used. 
   Binaries widen the confidence intervals towards low mass segregation values, 
   thereby preventing a firm conclusion as to the presence of an IMBH (see Section~\ref{compa}).} 
   \label{keyplot} \end{figure}

Figure~\ref{keyplot} (left panel) shows the M\,10 mass segregation
profile (red line) superimposed to the green and blue shaded areas
corresponding to the $2\,\sigma$ confidence regions defined by the
simulations without an IMBH and with an IMBH of mass $1\%$ of the total
cluster mass respectively. The observational profile lies within the
superposition of the two shaded areas. Therefore, while it is possible
that an IMBH is lurking in the center of M\,10,  we cannot rule out
alternative explanations. In particular, the lower envelope of the green
shaded area is defined, towards the center, by runs containing
primordial binaries with an initial binary fraction of $10\%$. This is
expected, because binaries act as an energy source capable of halting
core collapse, as shown by \cite{tre07b, tre07c} and consequently can
partially mimick the behavior of an IMBH, so they give rise to a
shallower profile with respect to runs containing only single stars. The
right panel of Figure~\ref{keyplot} shows the same plot of the left one,
where only the $32$k-particle simulations with no primordial binaries
were included (binaries do form dynamically in these runs, but only in
small numbers). In this case the data would be incompatible with the
no-IMBH scenario, and we would conclude that an IMBH is likely present
in the core of M\,10.

\begin{figure} \centering
   \includegraphics[width=0.49\textwidth]{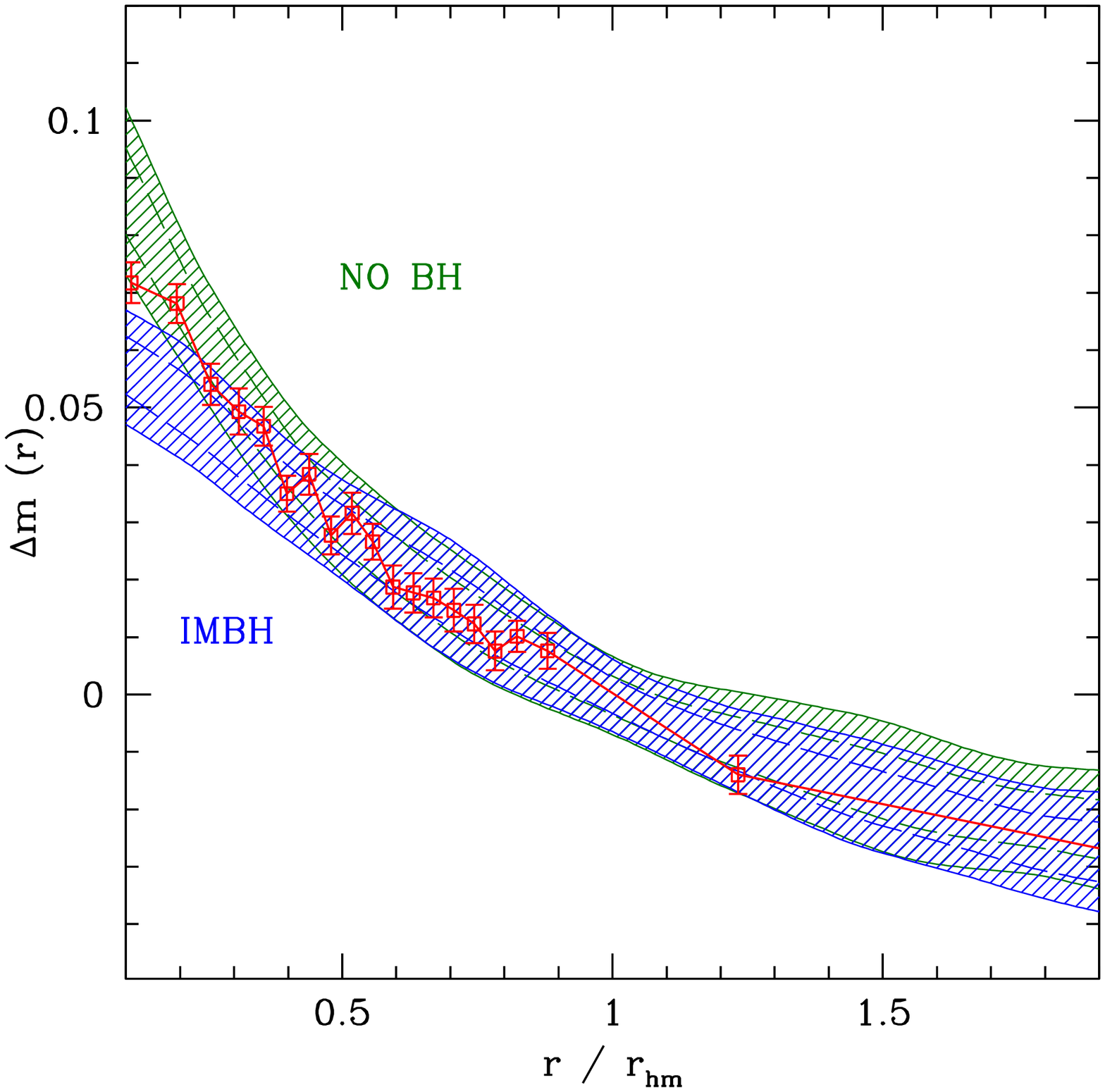}
   \includegraphics[width=0.49\textwidth]{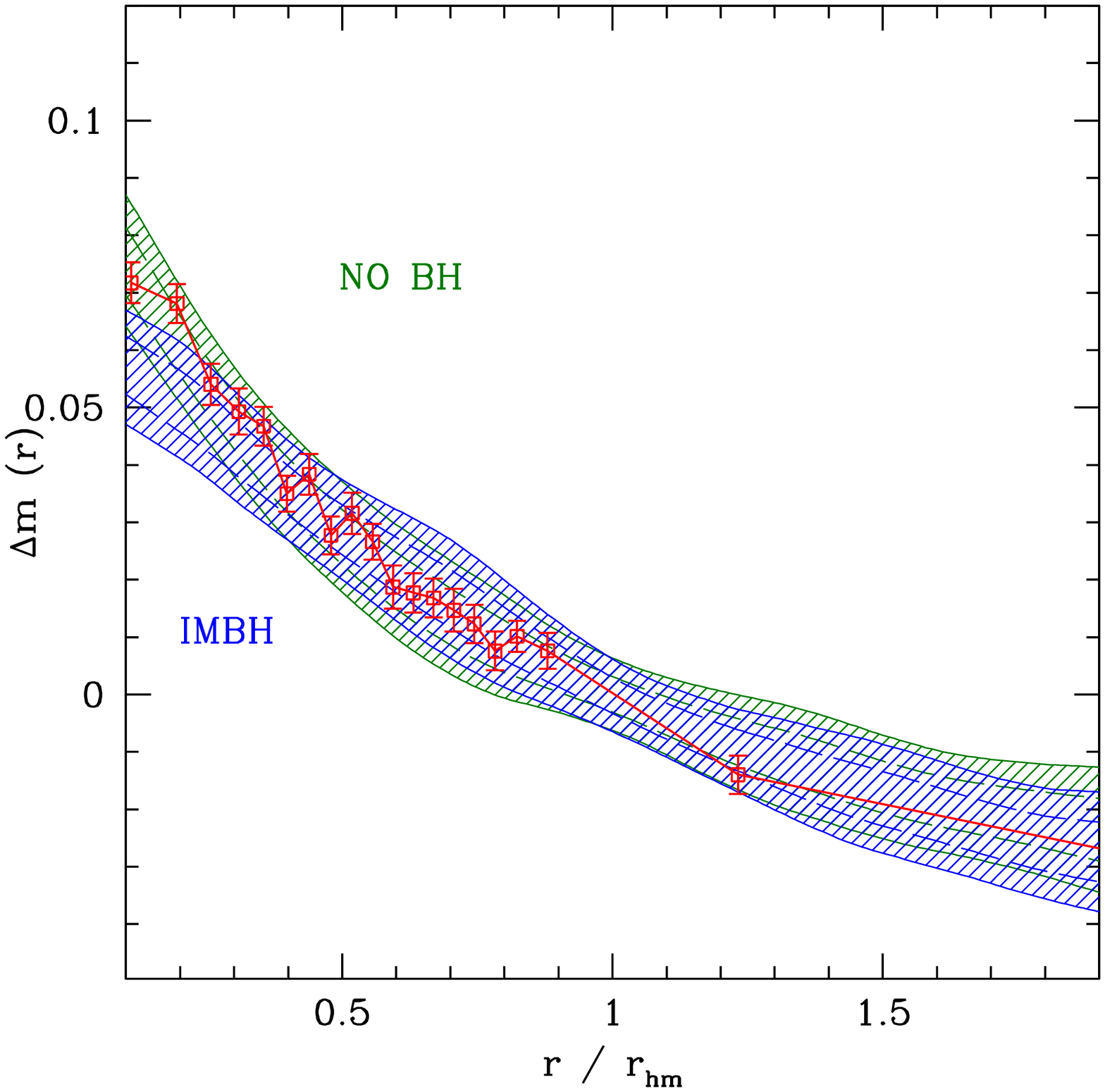}
   \caption{the same kind of plot as shown on Figure~\ref{keyplot}. Here
   we included simulation with a 3\% and 5\% of binaries on the left and
   right panel respectively.} \label{key_bin} \end{figure}

The two plot described above clearly show the crucial role played by
binary stars in the dynamical evolution of the cluster. In order to
investigate in more detail the effect of the binary fraction in the
predictive capability of our model, we compare the observed mass
segregation profile of M\,10 with the prediction of the $32$k-particle
IMBH-free simulations with $3\%$ and $5\%$ primordial binary fraction
(respectively left and right panel in Figure~\ref{key_bin}) .  It is
apparent that these simulations can reproduce the observations fairly
well, i.e. that in the absence of an IMBH a binary fraction in this
range is required to explain  the observations of mass segregation in
M\,10.

We therefore conclude that a precise measurement of the MS binary
fraction in M\,10 would allow us to reach a firmer conclusion on the
presence of an IMBH. A dynamically significant fraction of binaries is a
viable alternative to the IMBH hypothesis for explaining the observed
mass segregation profile, since both can be active energy sources
capable of halting core collapse and quenching mass segregation.

\section{Summary and conclusions}
\label{Conc}

In this work we present a study of the dynamical state of the globular
cluster M\,10 through the characterization of the radial properties of
its stellar LF and MF. We used a combination of archival deep ACS and
WFPC2 images to sample cluster stars from the core regions through to
$\sim 2.5\,r_h$ in the mass range $0.25 - 0.8$\,M$_{\odot}$. The full
data set was divided in six regions at different distances from the
cluster center and characterized by the same photometric completeness. 
The LF was calculated for each region and converted to a MF using the
mass--luminosity relationship of \citet{ba97} for the metallicity of the
cluster. Each of the local MFs was fitted using a simple power law  of
the type $dN/dm \propto m^\alpha$ and the best fitting indices were
found to be $\alpha=0.7, 0.4, 0.1, -0.3, -0.6, -0.9$ from the center
outwards. A positive value of $\alpha$ means that the number of stars
decreases with decreasing  mass. This radial change in the MF index is a
clear sign of mass segregation in M\,10. To better characterise this
effect, we built a  multi-mass Michie--King model to reproduce the
observed radial variation of the MF and the cluster's SBP and veolcity
dispersion and found that  the distribution of the stars in the cluster
is  compatible with a condition of equipartition of energy. The GMF as 
obtained by the Michie--King model is a power-law with index
$\alpha=-0.7$, in agreement with the value $\alpha\sim-0.6$ near the half
mass radius ($r_{hm} = {124}^{\prime\prime}$). 

In order to investigate in more detail the dynamical history of M\,10,
we used N-body simulations that include the presence and evolution of
binaries, with the goal to understand whether the current mass
segregation profile requires the presence of an IMBH. Within the current
uncertainties, we cannot exclude an IMBH of mass $\sim 1\,\%$ of the
total cluster mass, although a plausible alternative explanation of the
observed mass segregation profile is the presence of a substantial
amount of binaries, likely primordial in  origin, capable of
significant dynamical effects. We show that N-body simulations
initialized with a primordial binary fraction in the range of $3-5\,\%$
are effectively able to correctly predict the observed mass segregation
profile of M\,10.

M\,10 is a better candidate than NGC\,2298 for an observational
follow-up aimed at detecting an IMBH. While the latter cluster does not
contain an IMBH according to \cite{pas09}, M\,10 shows a shallow mass
segregation profile which could be interpreted as the fingerprint of an
IMBH.  The comparison of the observed mass segregation with the
simulations clearly demonstrates that the cluster binary stars play a
key role in defining its dynamical state and therefore cannot be
ignored. The binary fraction of the cluster needs to be better
constrained.

\acknowledgments

We thank the anonymous referee for the useful suggestions 
and comments that helped to signiÞcantly improve the overall quality of the paper .
Part of this research was carried out while Giacomo Beccari, Mario
Pasquato, Guido De Marchi and Michele Trenti were participating to the
Kavli Institute for Theoretical Physics (UCSB, Santa Barbara,
California) program on "Formation and evolution of Globular Clusters".
Support for grant HST-AR-11284 was provided by NASA through a grant from
STScI, which is operated by AURA, Inc., under NASA contract NAS 5-26555.
This work was supported in part through NASA ATFP grant NNX08AH29G and
by the National Science Foundation under Grant No. PHY05-51164.

{\it Facilities:} \facility{HST (ACS), JKT}.

\clearpage




\end{document}